# Soloist: Generating Mixed-Initiative Tutorials from Existing Guitar Instructional Videos Through Audio Processing

BRYAN WANG

University of Toronto, Ontario, Canada, bryanw@dgp.toronto.edu

MENGYU YANG

University of Toronto, Ontario, Canada, my.yang@mail.utoronto.ca

TOVI GROSSMAN

University of Toronto, Ontario, Canada, tovi@dgp.toronto.edux

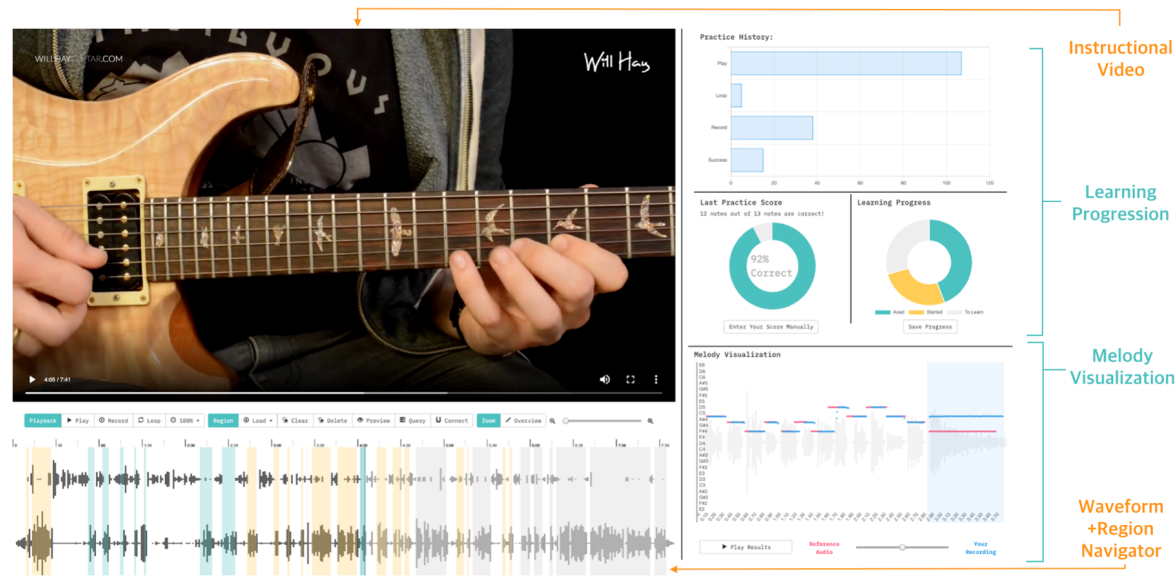

Figure 1. The Soloist user interface. Soloist performs audio processing on raw videos (Instructional Video) to provide navigation aids (Waveform + Region navigator) and real-time feedback (Melody Visualization, Learning Progression).

Learning musical instruments using online instructional videos has become increasingly prevalent. However, pre-recorded videos lack the instantaneous feedback and personal tailoring that human tutors provide. In addition, existing video navigations are not optimized for instrument learning, making the learning experience encumbered. Guided by our formative interviews with guitar players and prior literature, we designed Soloist, a mixed-initiative learning framework that automatically generates customizable curriculums from off-the-shelf guitar video lessons. Soloist takes raw videos as input and leverages deep-learning based audio processing to extract musical information. This back-end processing is used to provide an interactive visualization to support effective video navigation and real-time feedback on the user's performance, creating a guided learning experience. We demonstrate the capabilities and specific use-cases of Soloist within the domain of learning electric guitar solos using instructional YouTube videos. A remote user study, conducted to gather feedback from guitar players, shows encouraging results as the users unanimously preferred learning with Soloist over unconverted instructional videos.

CCS CONCEPTS • **Human-centered computing~Human computer interaction (HCI)~Interaction paradigms~Graphical user interfaces**

**ACM Reference Format:**
Bryan Wang, Mengyu Yang, Tovi Grossman. 2021. Soloist: Generating Mixed-Initiative Tutorials from Existing Guitar Instructional Videos Through Audio Processing. In *CHI Conference on Human Factors in Computing Systems (CHI '21), May 8–13, 2021, Yokohama, Japan. ACM, New York, NY, USA.* 22 pages. https://doi.org/10.1145/3411764.3445162

## 1 INTRODUCTION

The proliferation of online instructional videos has changed the way people learn skills [33, 34], and music is no exception [43]. Online video lessons for different musical instruments receive millions of views, with guitar being one of the most popular. Seeing this trend, stakeholders from traditional settings such as music institutions [69] and guitar manufacturers [70] have also started curating online video lessons on MOOCs [71, 72] or their own platforms [69, 70].

There are several advantages to learning guitar with online instructional videos. They can be accessed anytime and anywhere, and the rich variety of available videos grants users the freedom to identify specific content they wish to learn. However, browsing music instructional videos while practicing instruments can be difficult and encumbering [43]. A recent study [43] revealed the need for users to be able to select and play a specified interval in a video repeatedly. In another study, ~78% of the interactions made to navigate music instructional videos were to replay certain parts of the video, likely due to the repetitive nature of music practice [9].

Another widely recognized difficulty of learning music with pre-recorded videos is the lack of immediate feedback and personalized tailoring that human teaching affords [20, 42]. While intelligent tutoring systems (ITS) [21, 22, 62, 65] and interactive tutorials [1, 23, 24, 39] promise to bridge this gap, the domain of music is particularly challenging since it requires additional efforts to process external audio signals and recognize what the user is playing on their musical instrument. Furthermore, existing training systems often require the handcrafting of new tutorials from scratch [44, 66], making it prohibitively time-consuming and costly to customize tutorials for individual users.

In this paper, we present Soloist, a mixed-initiative guitar solo training system that automatically transforms existing music instructional videos into interactive tutorials. Soloist leverages deep-learning based audio processing to extract musical information and demonstration moments from raw video without requiring any prior knowledge of the video. Soloist automatically segments the video into regions containing instrument demonstrations and provides interactive visualizations that help users efficiently identify and navigate to desired video segments. Soloist also records user performance and then compares it with the corresponding audio from the instructional video to provide various immediate feedback, including melody visualization, note correctness score, and learning progression.

We first present a formative interview to understand the experiences of guitar players with existing online video lessons. Guided by the interview and literature, we introduce a set of design considerations for the Soloist framework. We demonstrate the capabilities and specific use-cases of Soloist within the context of learning electric guitar solos with instructional YouTube videos. We then present the interface and backend implementation in detail. After that, we present a technical evaluation that shows that Soloist's video segmentation outperforms the other two baselines and achieves reasonably consistent performance with human judgment. Finally, we present the results from a remote user study with eight guitar players. The results show that users unanimously preferred learning with Soloist over traditional instructional videos and point to the framework's benefits and future potential. Taken together, our work offers the following contributions:

1. An automatic pipeline to generate tutorials from existing music instructional videos using audio processing.
2. A set of video navigation tools addressing the limitations of learning music using traditional videos.
3. The design and implementation of Soloist and a preliminary evaluation with hobbyist guitar players.

## 2 RELATED WORK

Our work builds upon research in intelligent tutoring systems, interactive video tutorials, and music learning systems.

### 2.1 Intelligent Tutoring Systems

There has been a decades-long pursuit of using artificial intelligence techniques to replicate learning experiences with human tutors, namely the intelligent tutoring system (ITS) [8, 60, 62]. The principle is to design computer programs able to recognize the user's learning activities and provide personalized feedback to help users correct errors. While the original goal is to replicate human teaching, it has been shown that ITSs can sometimes be more effective than human tutors [35, 52]. Some intelligent tutoring systems such as Duolingo have become successful commercial products with impressive learning outcomes [63]. Closely relevant to our work, researchers have also investigated ITSs for music theory education [5, 61]. For example, Maestoso [61] helps novices learn music theory through sketching practice of quizzed music structures. Our system builds upon prior work of intelligent tutoring systems and contributes a general approach to transforming existing music instructional videos into intelligent tutoring systems for instrument learning.

### 2.2 Interactive Video Tutorials

Interactive video tutorials have been studied intensively for various domains, such as software learning [1, 4, 11, 23, 24, 29, 39, 49, 53] or physical activity [12, 25, 38]. Prior work has investigated both the facilitation for authors to generate interactive video tutorials [11, 12, 25, 51] and for users to effectively follow the video instructions [1, 24, 53]. Although useful, video tutorials are hard to navigate [1, 33, 51]. Prior work has leveraged video segmentation to split an existing video into conceptual chunks to help people search for information [1, 11, 12, 51]. In terms of the segmentation method, most closely relevant to our work is Waken [4], which reverse engineers an input software tutorial video to recognize UI components using computer vision techniques. Soloist is a close counterpart for music instructional videos, except we use audio signal processing to extract the demonstration moments and musical notes from raw videos. To navigate a segmented video, prior work has leveraged interactive timeline markers [1, 4, 24, 29, 33, 34], thumbnail images [4, 11, 23, 51, 53], transcript text [51], and clickable elements overlaid on the video [49]. Soloist's video navigation also falls into the above categories but is designed in an audio-oriented manner, providing a waveform timeline with clickable regions to replay segments.

### 2.3 Music Learning Systems

Music has been traditionally taught in a "master-apprentice" relationship where the student receives personalized and proper instruction from an experienced individual [20]. However, human tutors are not always available and can be expensive. Prior work has explored how computers can be harnessed to develop music learning systems. In early works,

Dannenberg et al. [14] used a hand-designed expert system to evaluate user performance and recommend corresponding remedial lessons. Similar to Soloist, Strummer [44] is an interactive system for guitar practice that focuses on teaching chord strumming while we focus on guitar solos. The BACh [67] system measures user cognitive workload with brain sensing and adjusts the difficulty of the lesson accordingly. Commercial tools are also available for teaching musical instruments. For example, Yousician [73] is an online platform that gamifies instrument learning with a scrolling fretboard and provides real-time feedback. However, existing music learning systems often run into limitations where the curriculum is predesigned, thus limiting the amount of teaching resources available [14, 44, 67]. In contrast, the Soloist system adapts to existing video lessons and automatically generates personalized tutorials.

Past work has also leveraged various sensory signals to facilitate music learning [28, 31, 42]. For instance, EMGuitar [31] uses electromyography to detect fine-grained hand and finger positioning for guitar. Our system minimizes the use of additional sensors and leverages only acoustic data that can be easily obtained using computer microphones or recording interfaces.

## 3 FORMATIVE INTERVIEWS

To help formulate our system's design, we conducted formative interviews with eight electric guitar players to understand how players with different levels of expertise utilize online video lessons for learning guitar. We recruited interviewees who have prior experience in learning guitar using YouTube videos. We asked the participants to self-report their proficiency using the following criteria: Professional, if playing guitar is involved in their jobs; hobbyist, if they are not professional but experienced with playing guitar; novice, if they are not yet experienced guitar players. Three of them (P1-P3) self-reported as novices (< 2-year experiences), four (P4-P7) as hobbyist players (5 ~ 20-year experiences), and one (P8) works as a professional musician and guitar tutor. We conducted the semi-structured interviews over video conference, which lasted around half an hour. The interviewees received a 10 CAD compensation for participation. We started by asking interviewees about their general impression on learning guitar with YouTube and then focused on specific questions developed from prior literature to understand the efficiency and efficacy of learning guitar using YouTube videos. We summarize four key findings that were commonly mentioned by the interviewees below.

### 3.1 Learning Guitar with Online Videos is Fast but Needs Additional Initiative from Users

All interviewees described YouTube as a low-cost and convenient tool to quickly explore community-uploaded learning content. Moreover, video demonstrations contain significant amounts of visual and auditory details of guitar performances, making them easier to digest and learn from than guitar tabs or the original songs. However, interviewees also commented that the quality of video lessons varies between different creators. Sometimes, videos may contain incorrect information and require additional effort from the user to continue searching. Lastly, all three novices commented that they sometimes feel less motivated when learning with videos because *"YouTube won't supervise me like a tutor does."* (P3).

### 3.2 Learning Guitar with Videos Lacks Feedback for Improvement

The interviewees perceived feedback to be vital for learning guitar and commented that it is hard to objectively evaluate their own performances when learning with videos due to the absence of feedback. They have to examine their performances either by listening while playing or listening to a recording of their practice afterwards. Interestingly, we found that advanced players were more likely to record their practice and emphasize its importance. One advanced player (P4) said recordings reveal more details since he cannot focus equally on both playing and listening, *"Sometimes you thought you've played an 80, it's in fact a 60 when you listen to the recording."*

We also identified a set of typical mistakes that guitar players might make when learning guitar solos. We found that common mistakes for novice players are mostly technique related, e.g. unstable tempo, wrong notes. Moreover, it is common that novices' mistakes went unnoticed when they practiced alone, implying the need for additional notification of mistakes. On the other hand, advanced players tend to make more conceptual mistakes, such as imperfect musical expressions or interpretations. All groups of interviewees expressed the desire to obtain feedback for correcting their mistakes but confirmed that videos barely provide any, as the information only propagates unidirectionally.

### 3.3 Existing Video Navigation is not Optimized for Instrument Learning

All interviewees mentioned that they sometimes felt inconvenienced when learning with videos because existing navigations are not optimized for learning guitar. These inconveniences mostly fall under one of the following two categories: (1) *Frequent attention transitions between guitar and video player*: For instance, to practice a particular part of a song, users need to repeatedly move their hands off the instrument to control the video. This was common for all interviewee groups, further extending prior findings [43] to non-professional musicians. (2) *Difficulty in precisely selecting the timestamp for playback*: For example, it is challenging to locate specific pieces of content or notes on the timeline. These operations are particularly crucial for instrument learning because unlike videos in other domains, a great deal of information can be contained within a small length of time (e.g., multiple fast notes).

### 3.4 Novices Focus on Building Muscle Memory while Advanced Players Also Look for Inspiration

Finally, we observed a novice-to-expert transition in watching behavior. We found that novices mainly watch tutorial-style videos that demonstrate step by step instructions on playing a song or a technique. This is because their primary goal is to build muscle memory. The advanced players commented that they would also utilize tutorial-style videos when they need to learn how to play a new song quickly, e.g., practice for upcoming rehearsals. However, with more guitar expertise, they gradually switch to using online videos to draw inspiration or learn music theory. For example, they would watch video lessons where famous guitarists explain how they create a song or share their practice tips.

## 4 DESIGN GOALS

Guided by both our formative interviews with guitar players and prior literature, we established five design goals to guide the creation of the Soloist system. Note that Soloist is designed to address the intrinsic limitations of guitar learning experiences obtained from online instructional videos and is not meant to serve as a replacement to traditional 1-1 human instruction.

**D1: Utilizing Existing Instructional Videos.** There is already a wealth of instructional videos for musical training on the Internet [64]. Therefore, instead of handcrafting new tutorials from scratch, it would be useful to be able to extract musical information from existing videos and convert them into interactive tutorials.

**D2: Support Efficient Navigation within Videos.** Soloist should support efficient video navigation to improve the user's learning experience. For example, the system should expedite repetitive practice [16, 43] while minimizing transitions between playing the instrument and navigating the video [43]. Additionally, users should be able to easily discover sections within the video that they are interested in [1].

**D3: Provide Musical Feedback and Guidance.** Soloist should be able to provide individualized diagnoses of errors and immediate informative feedback [16] based on the user's performance to help improve their playing. We focus on providing feedback to correct technique errors which can be evaluated quantitatively.

**D4: Track and Inform User's Learning Progress.** Soloist should clearly present the user's progress whenever some milestone is reached. This helps reinforce the user's motivations to keep learning [48].

**D5: Mixed Initiative Tutoring.** Soloist is built upon ITS's core concept of using artificial intelligence to assist learning, which will inevitably introduce uncertainties. To balance automation and controllability [56], a mixed-initiative approach should be used to include humans for error correction.

## 5 SOLOIST FRAMEWORK

To design a system which fulfills these design goals, we propose a full-stack framework consisting of a backend server and a frontend web interface (Figure 2). To create a learning session, the user selects from a rich source of videos from YouTube (D1). The audio from the video is preprocessed to obtain the necessary information. That information, along with the video, is presented on the frontend web interface. During a learning session, any computationally demanding task, such as processing the user's recording, is sent to the backend server. We now discuss the Soloist system interface followed by its implementation details.

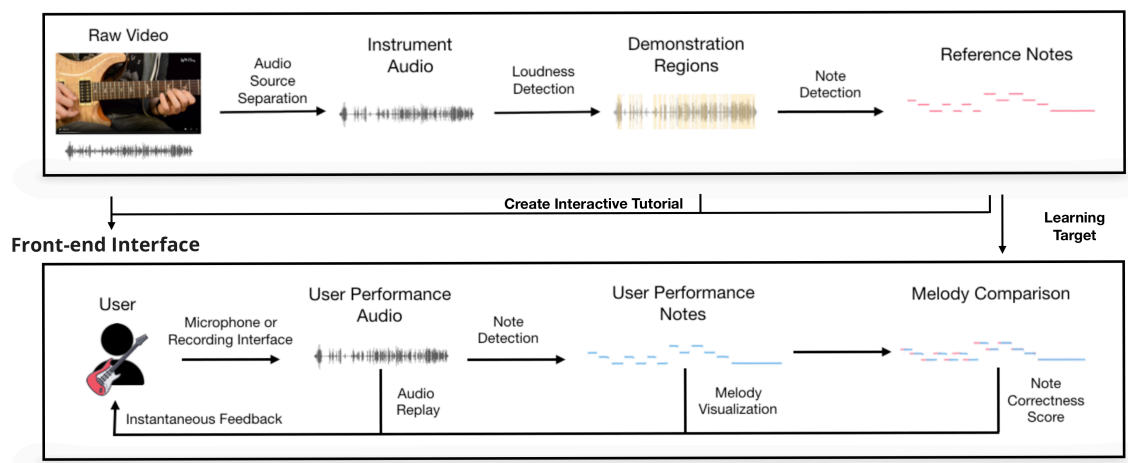

Figure 2. The Soloist framework consists of a raw video, a backend processing pipeline, and a frontend web interface.

### 5.1 Soloist Interface

The full view of the Soloist system interface is illustrated in Figure 1. We now describe each component in detail.

### 5.2 Instructional Video

Soloist takes an off-the-shelf video tutorial as input and augments it to provide an interactive intelligent tutoring experience. We consider videos that contain a tutor who teaches electric guitar solos with an electric guitar. An electric guitar solo is a melodic passage, usually monophonic, written for an electric guitar. Compared to other types of guitar performance, the electric guitar solo tends to contain virtuosic techniques and varying degrees of improvisation, making it often the most significant instrumental section in rock and related music genres.

### 5.3 Waveform and Region Navigation

Navigating music instructional videos is challenging as they normally do not contain explicit steps [9] and requires transitions of attention between the instrument and video player, making learning encumbered [43]. Visual-based navigation methods (e.g., thumbnail previews [6], scrubbing [45, 46]) are less useful for audio-based music instructional videos. We introduce the design of *waveform+region* navigation, a standard interaction paradigm used in professional music production tools [74, 75] to facilitate music instructional video navigation (D2). While similar designs have been leveraged in HCI literature for music-related interfaces [57, 58], none of them were used to explicitly address the challenges of navigating music instructional videos.

*5.3.1 Waveform Player*

The video navigator of Soloist provides an interactive waveform timeline which visualizes the video lesson's audio (Figure 3). The timeline consists of two rows of waveforms stacked on top of each other. The top row visualizes the amplitudes of the voice track, taken from the narration of the instructor in the video. The bottom row shows the waveform of the musical instrument. When the video is being played, a playback head moves across the waveforms. Clicking on the waveform timeline will skip the playback head to the clicked position. Users can also use the *ZoomSlider* on the toolbar to zoom in and out of the waveform or press the *Overview* button to instantly reset the zoom level. Displaying the separated waveforms helps users locate specific regions where the instructor is speaking or where the instrument is playing.

*5.3.2 Regions*

To further facilitate navigation and rehearsal, regions, which represent semantic chunks of playback, are visualized as rectangles spanning across a specified interval. Region-based playback is especially useful because music is built upon segmented structures such as phrases and bars. Regions can be created automatically by the system or specified manually by the user (D5). To create a region, the user drags horizontally on the waveforms with a mouse. Once a region is created, the user can click on it and the waveform player will play the interval specified by the region. The *Loop* button can be used to play the region repeatedly. The user can drag the boundaries to refine the start and end of a region and also freely switch between region-based playback and continuous playback.

*5.3.3 Navigational Toolbar*

Soloist provides a set of additional navigation tools within three groups of buttons: Playback, Region, and Zoom. We designed the navigation tools to expedite repetitive practice that helps users establish muscle memory. In the Playback group, the user can click Play to begin playback, Loop to play a region repeatedly, and Record to rehearse the current region. The user can change the playback speed of the video by selecting a value in the Speed dropdown. In the Region group, users can Load all the pre-segmented regions. They can also Delete a region or Clear all presented regions. Clicking the Preview button and hovering the cursor on different regions will display the melodic lines being played in the associated regions. Moreover, Soloist supports region Query, where users can create or select an existing region and display all other regions containing a similar melody. Finally, the Connect button plays all the displayed regions consecutively, which helps users practice multiple phrases in a row. The designs of Query and Connect were motivated by prior work in video learning [24, 33] which explored features that allow users to find all instances of a specific query to aid in learning and navigation. We hope to see if such techniques would help users practice queried musical techniques or phrases that appear multiple times throughout the video.

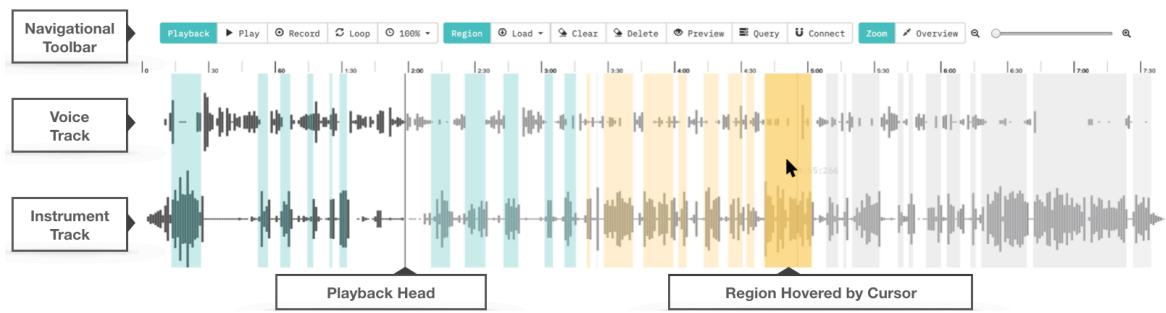

Figure 3. The waveform + region navigator, consisting of a navigation tool bar and an interactive waveform timeline with regions. The audio is separated into voice track and instrument track. Colors of regions represent different practice states. Cursor hovering will highlight regions.

**5.4 Rehearsal Recording and Feedback**

Our goal is to design a system that turns any instructional video (D1) into an interactive tutorial that can provide real-time feedback (D3). To this end, the segmented regions are considered as customizable practice curriculums, based on principles that have been shown to be effective for music learning [13, 27] and skill acquisition in general [16, 47]. We also utilize some principles of gamification, such as score and progression, to motivate the learner to complete the curriculum progressively [48].

*5.4.1 Recording*

Soloist provides instantaneous feedback (D3) by analysing the recording of the user's performance. The user first selects a region they want to learn and clicks the *Record* button. Soloist will then play the region for the user to follow along while recording the audio of the user's performance. Musical note detection is then performed on the recording and reference audio from the selected region to provide three types of instantaneous feedback: Melody Visualization, Audio Replay, and Note Correctness Score. Recording can be done at any playback speed.

*5.4.2 Melody Visualization*

To provide visual feedback on user performance, Soloist visualizes the melodic lines of both the user's recording and the instructor's demonstration (Figure 4A). With this visualization, users can quickly identify deviations from the reference performance such as incorrect pitch or temporal shifts. For instance, when practicing a new scale, users might not notice that they incorrectly played one of the notes higher by a halftone, which is typical for novices who have not developed a good sense of pitch (Figure 4B).

Moreover, since many guitar playing techniques, such as string bending or vibrato, result in pitch shifts, users can check whether they have performed a technique correctly. For example, it is quite common for novices to be unable to detect the existence of vibrato by ear. With melody visualization, they can see vibratos represented as continual fluctuations (Figure 4C). Users can also identify if they failed to bend strings to the expected pitch (Figure 4D).

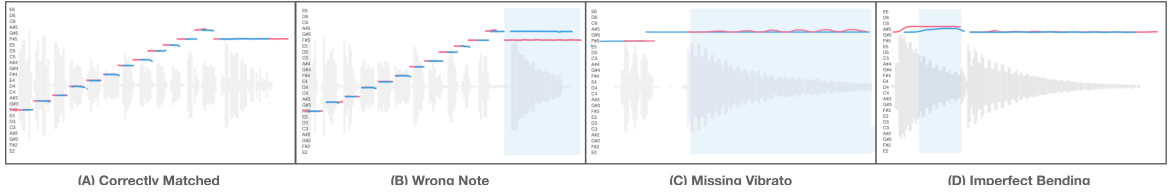

(A) Correctly Matched (B) Wrong Note (C) Missing Vibrato (D) Imperfect Bending

Figure 4. Example cases where Melody Visualization can help users self-assess their practices. The red lines are the reference audio from the video and the blue lines represent the user's performance. (A) An example of two melodic lines matched correctly. (B)-(D) are mistakes that can be hard for novice to notice by ear but can be identified with visuals. (B) An incorrect note in a note sequence. (C) User did not perform vibrato (D) User failed to bend the string to the correct pitch.

### 5.4.3 Audio Replay

Soloist provides auditory feedback by encouraging users to listen to their recorded performance and compare it with the tutor's instruction [27]. As shown in Figure 5A and Figure 5B, Soloist displays another waveform player overlaid by the melody visualization. This allows users to listen to both their performance and the reference simultaneously by clicking the *Play Results* button. They can also select regions to only play a specified interval. Soloist also automatically highlights regions containing potential mistakes by identifying inconsistencies among the two note sequences. During playback, a mixing slider allows the user to adjust the relative volumes of the reference audio and their own recording. They can choose to listen to only one of the tracks or they can adjust the slider to listen to a mixture of both.

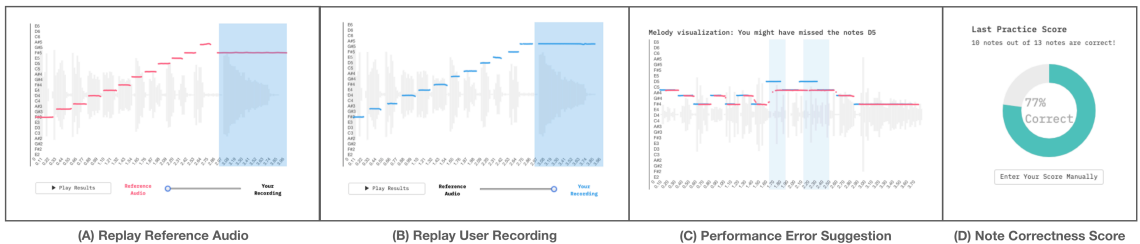

(A) Replay Reference Audio (B) Replay User Recording (C) Performance Error Suggestion (D) Note Correctness Score

Figure 5. (A-B) Audio player overlaid with the melody visualization. Clicking Play Results would play both the user performance and the reference simultaneously. Dragging the mixing slider adjusts the relative volumes and opacity of both melodic lines. (C-D) Soloist displays how many notes the user played correctly (right) and the name of the notes missed (left). The blue regions on the melody visualization suggest moments that might contain errors.

### 5.4.4 Note Correctness Score

The last form of recording feedback is a score ranging from 0% to 100% indicating the correctness of the user's performance (D3) along with the number of correctly played notes and names of missed notes, as shown in Figure 5C and Figure 5D. However, inaccuracies are inevitable in any machine learning model. Therefore, the system allows users to override the scores given to them (D5) by clicking on the *Enter the Score Manually* button. A text field will appear for the user to enter their self-assessed score (Figure 6A). We decided not to provide score feedback on temporal aspects of the music since tutors often break down phrases and demonstrate in free rhythm without following a consistent tempo.

As a result, minimizing temporal differences between user performance and tutor demonstration may not be as meaningful as pitch.

## 5.5 Learning Progression Feedback

To improve progress feedback (D4), Soloist color-codes each region in the timeline. The colors grey, yellow, and teal represent the stages *To Learn*, *Started*, and *Aced*, respectively (Figure 3). When a region is first created, it will be labelled grey as a *To Learn* region. Once the user clicks on that region or the playback head enters the region, it will be marked yellow as a *Started* region. The region will then turn teal into an *Aced* region after the user obtains a 100% note correctness score for that region. Feedback on progress across the entire video is also provided. A doughnut chart with three sub-divisions represents the proportion of regions in each of the three learning stages (Figure 6B). The breakdown of the practice history is also presented to the users in the form of a bar chart, which illustrates how many times the user played, looped, recorded, and aced each region, as shown in Figure 6C.

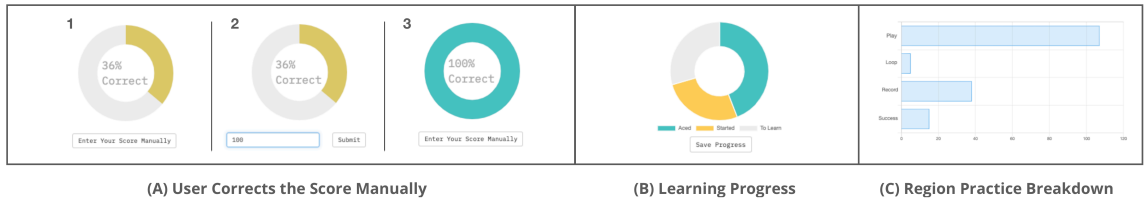

(A) User Corrects the Score Manually  (B) Learning Progress  (C) Region Practice Breakdown

Figure 6. (A) The process to override the score given by the system. 1. Click "Enter Your Score Manually". 2. Enter self-assessed score and click "Submit". 3. The score would then be updated. (B) A doughnut chart with three sub-divisions represents the proportion of regions in each of the three learning stages. (C) The breakdown of the practice history.

## 6 SYSTEM IMPLEMENTATION

The Soloist system is a web application implemented in vanilla JavaScript. The frontend interactive visualizations are developed with Wavesurfer.js and Chart.js. The backend processing is written in Python running on a Flask server. We developed the system on a MacOS machine and deployed it on a virtual machine instance with a Nvidia Tesla K80 GPU on the Google Cloud Platform running an Ubuntu environment. We use CREPE [32] for fundamental frequency detection and Spleeter [55] for audio source separation, both based on TensorFlow.

### 6.1 Audio Signal Processing Modules

Extracting musical information from a guitar performance requires complicated audio signal processing. To accomplish this, Soloist leverages two deep-learning based audio signal processing modules: audio source separation [3, 41, 50] and fundamental frequency (f0) estimation [7, 10, 32].

#### 6.1.1 Audio Source Separation

Audio source separation splits a mixed music track into multiple source tracks, or stems, such as voice, drums, and bass. Our framework leverages this to separate the audio of the instructional videos into voice and instrument tracks. We use Spleeter [55], an open-source Python package providing state-of-the-art pre-trained models that can separate an audio faster than real-time on a single GPU.

*6.1.2 Fundamental Frequency Estimation*

Fundamental frequency (f0) estimation takes an audio signal as input and estimates the f0 of each frame. F0 is defined as the lowest frequency of a periodic waveform and can be used to estimate the pitch. F0 detection can be performed on either monophonic music, consisting of a single line of melody (single-f0) [10, 32, 47], or polyphonic music, consisting of two or more simultaneous lines of independent melody (multi-f0) [7, 59].

In our framework, we estimate the notes played by both the user as well as the instructor in the video using CREPE [32], a state-of-the-art monophonic pitch tracker based on a deep convolutional neural network. Along with the estimated f0, CREPE also calculates the estimation confidence of each time frame (frame size = 0.1s). We chose single-f0 estimation over mulit-f0 estimation to maximize estimation accuracy since multi-f0 estimation remains an ongoing research problem [7].

## 6.2 Backend Processing

*6.2.1 Video Segmentation*

To facilitate video navigation (D2), Soloist automatically segments the video into regions where either the tutor is explaining concepts verbally or is demonstrating on the instrument. To achieve this, a potential approach is to perform sound classification [26] to detect whether each frame within the video contains speech or instrument and aggregate those frames. However, training a dedicated classification model for this task requires labelled datasets with frame-level annotations. Unfortunately, most datasets [19, 30] for audio classification are labelled by excerpts of >1s in length which may not generalize well to a smaller granularity. Moreover, classifying the sound of an instrument unseen by the model requires re-training with additional labelled data, which further hampers the generalizability of this approach.

Instead, we propose a pipeline that does not require specific tuning. Our approach is motivated by an observation on music instructional videos: The audio mostly consists of only the tutor's voice and the instrument's sound. Therefore, by separating the voice and instrument track, we can reformulate the classification problem into a much simpler problem of detecting non-silent regions in both tracks. We first use Spleeter's 2-stem audio separation to obtain the voice and instrument track. Then, we use a similar approach as in DemoCut [12] to label non-silent sections across the waveform by analysing the loudness of each window for both tracks.

To calculate loudness, we first normalize the audio and use a 0.02s (n = 441) sliding window to calculate the root mean square (RMS) energy of each window. The RMS energy for a window of size $n$ is calculated by the equation:

$$RMS = \sqrt{(\sum_n X_i^2)/n} \;,$$

where $X_i$ is the amplitude value of the $i$th audio sample in the window. We then label each window as silent or non-silent based on a variable threshold ε. For each video, the threshold ε is obtained by finding the minimum point in the second order derivative of the Gaussian-filter-smoothed histogram of the RMS energy of its audio. Once every window has been labelled, we group the consecutive non-silent windows together as regions with a 2s threshold, i.e. if two notes are more than 2s apart, they belong to different regions. Regions with durations <1s are removed from the final results. The thresholds were chosen based on results of initial experiments which maximized accuracy from inspection.

*6.2.2 Musical Note Detection*

After obtaining the regions for the vocal and instrumental tracks, Soloist estimates the sequence of musical notes, i.e. melody, for all the instrumental regions using f0 estimation. This gives us a sequence of all the notes that have been demonstrated within the video, which we consider to be learning targets or playing references for the user. Soloist also conducts musical note detection to identify the notes the user played.

The complete procedure for Soloist's note detection during video pre-processing is as follows:

1. Take the separated instrumental track as input.
2. Perform f0 estimation on the input track.
3. Remove all estimations with <70% confidence.
4. Convert the remaining f0's to MIDI numbers.

We set a threshold of 70% for CREPE's estimation confidence to remove any noise from the output that may have been classified as notes. We use the equation:

$$M = 12 * \log_2 \left(\frac{F}{440}\right) + 69 ,$$

to convert f0 values into musical notes, which are commonly represented as MIDI numbers ranging from 0-128. Here, *F* and *M* stand for the input frequency and the output MIDI numbers, respectively. We round *M* and aggregate the consecutive frames with identical MIDI numbers into one note. We also keep a copy of the unrounded values to preserve small fluctuations or note transitions produced by playing techniques such as vibrato or string bending. The rounded numbers were used for melody comparison while the unrounded ones were visualized and presented to the user.

## 6.3 Correctness Score Calculation

The user's notes and the demonstrated notes are combined to assess the user's performance. Given a target note sequence, *T*, and a note sequence from the user's recording, *R*, the note correctness score is calculated as follows:

$$Score = \frac{Length\ of\ LCS(T,R)}{Length\ of\ T} \times 100\%.$$

Here, LCS stands for longest common subsequence and the notes in both *T* and *R* are rounded MIDI notes. The note correctness score is proportional to how many notes in *T* are played in *R*. Specific sections containing potential mistakes are identified using Python's difflib, which is built upon Ratcliff and Metzener's string-matching algorithm [54].

## 6.4 Region Querying

Another feature that leverages the calculation of the correctness score is region querying, which allows the user to navigate to regions with similar melodies. We use the same formula as above to calculate scores between regions, but *T* becomes the query region and *R* is each of the remaining regions. Therefore, any sequence that fully contains the query would score 100%. To tolerate potential inaccuracies of musical note detection, regions that score >80% are regarded as containing the query sequence.

## 7 TECHNICAL EVALUATION

We conducted a technical evaluation of our video segmentation approach by comparing the results of the algorithm against those labelled by humans. Note that detailed evaluations for the audio source separation [55] and f0 estimation [32] can be found in their original papers.

### 7.1 Target Videos:

While we focus on electric guitar (EG) solo lessons in the paper, we were also interested in how well our approach can generalize to acoustic guitar (AG). As a result, ten each of the relevant top viewed AG and EG video lessons were downloaded for evaluation from YouTube. This gave us 1.5hrs of EG videos (min =4.8min, max = 13.3min) and 1.7hrs of AG videos (min = 4.4min, max = 16.3min).

### 7.2 Video Labelling:

Two of the paper authors conducted the labelling process, which involved highlighting regions within the video where the instrument was being played. One had 8 years of experience playing guitar while the other did not play guitar but had 15 years of experience with music. The set of videos was divided among the two labellers, with each labelling their own collection. They were told to generate the regions based on their musical judgment. The labelling interface consisted of a waveform-based video player similar to the one in Soloist, except that the audio was not separated into voice and instrument tracks.

### 7.3 Method:

We quantify how similar Soloist's segmentation is with human segmentation at both frame level and segment level. At the frame level, we divide each video into discrete timeframes of 0.02s each and calculate the precision, recall, and F1 score metrics using human segmentation as the ground truth. The timeframe window size is identical to the one used in the backend processing. At the segment level, we use boundary similarity [17], ranging between 0 and 1, where a score of 1 means the two segmentations are identical and a score of 0 means they are completely different. Since boundary similarity quantifies distances between boundaries in discrete intervals, we chose a granularity of 1s for how far apart in time two boundaries can be and still be considered the same. Similar to the approach in [18], we used a boundary edit distance window of 5 seconds (average length of a segment) as the maximum distance that two boundaries may span to be considered a near miss as opposed to a full miss. The penalty of a near miss is proportional to how far a pair of boundaries deviate within the near-miss threshold. Any pair of boundaries with a distance larger or equal to the near-miss threshold is penalized as a full miss. The algorithm's performance (algorithm) was also compared to two simple segmentation heuristics. The first randomly segments the video (random) while the second uniformly segments the video based on the average region length created by the human on the same video (uniform). These techniques were included to ensure our algorithm was performing better than what would be expected from random or arbitrary segmentations.

### 7.4 Results:

As shown in Table 1, the algorithm far outperforms the random and uniform segmentations across all metrics. The consistent scores for the frame-level metrics show that our algorithm can correctly predict whether to include each timeframe in a demonstration region or not. The boundary similarity scores representing segment-level correspondences are slightly lower than frame-level metrics for both AG and EG. This discrepancy is due to boundary similarity only considering each segment's start and end, regardless of how many correct predictions are within the segments. Nonetheless, the ~0.7 average boundary similarity score (AG and EG) indicates that our algorithm is reasonably accurate and better than other baselines.

Given the content's subjective nature, we did not expect perfect boundary similarity scores for our algorithm since there may be more than one "correct" way of segmenting a video [18]. For instance, interpretations can differ between when a region ends, and another begins. Our algorithm generated the regions based on a fixed time threshold of 2s while human labellers might group larger spaced notes together based on their musical judgments. We recognize that the algorithm's accuracy still has room for improvement, which motivates the mixed-initiative approach used in Soloist's design (D5). In the next section, we will explore the impacts of these benchmark accuracies on the end-user learning experience.

Interestingly, our algorithm's segmentations seem more accurate for AG. One possible reason is that for AG videos, most of the non-silent guitar regions were relatively long as they involve strumming repeated phrases, making it easier for both the algorithm and human labeller to identify them. On the other hand, the EG videos first demonstrate each of the individual notes involved before playing the whole phrase. Accordingly, the EG videos contain more short regions, causing additional discrepancies between the algorithm and the human labeller.

Table 1: The results of technical evaluation reported in Mean and SD, separated by acoustic and electric guitar videos.

|  | Algorithm-Human | Random-Human | Uniform-Human |
| --- | --- | --- | --- |
| **Acoustic Guitar** | | | |
| Precision | 0.949 (0.05) | 0.447 (0.21) | 0.459 (0.21) |
| Recall | 0.885 (0.12) | 0.465 (0.11) | 0.506 (0.03) |
| F1 Score | 0.911 (0.08) | 0.424 (0.15) | 0.453 (0.15) |
| Boundary Similarity | 0.731 (0.12) | 0.198 (0.05) | 0.227 (0.10) |
| **Electric Guitar** | | | |
| Precision | 0.947 (0.04) | 0.547 (0.11) | 0.524 (0.11) |
| Recall | 0.792 (0.13) | 0.521 (0.03) | 0.495 (0.02) |
| F1 Score | 0.855 (0.07) | 0.530 (0.06) | 0.505 (0.06) |
| Boundary Similarity | 0.661 (0.10) | 0.224 (0.02) | 0.391 (0.08) |

## 8 QUALITATIVE USER STUDY

We conducted a remote qualitative study to obtain initial observations and feedback on the design aspects and features of Soloist. The study was meant to collect high level subjective feedback on Soloist, not to formally evaluate its impact on long-term learning progression.

### 8.1 Participants

We recruited eight external participants from an online posting. We used the same criteria from the formative interviews to classify the participants' expertise. Four were self-reported novices on electric guitar (P1-P4) ranging from 2 weeks to 6 months of experience and the other four reported themselves as hobbyists (P5-P8) with 2 to 9 years of experience. All participants commented they have been using YouTube for nearly as long as they have been playing guitar. 7 of the 8 participants commented that they use YouTube quite often, while the remaining participant (P3) mentioned that he uses YouTube only when transcribing new songs. Desktop or laptop computers were reported as the primary device used to view YouTube videos. Participants were compensated 30 CAD for their participation.

### 8.2 Methodology

The study was conducted remotely. Participants used their own computer and guitar and communicated with the experimenter using a videoconferencing tool. The participants were asked to share their screen and computer sound while using Soloist so that the experimenter could record the study. To record their playing, six participants connected their guitar into the computer using audio recording interfaces while the other two used laptop microphones to record the output of a guitar amplifier.

Before the study, users were told to choose any YouTube solo lesson they wanted to learn. The chosen videos were processed using Soloist's back-end processing to ensure the system was ready for the study. The participant received a URL link to the system running on a Google Cloud virtual machine in a data center closest to their location to minimize network latency. While we encouraged the users to practice and turn all the regions into Aced regions, we explicitly made it clear that they could use whatever learning strategies they wanted to when practicing with Soloist. We did not expect them to finish the video in one session and they were not being tested for how well they learned the solo.

### 8.3 Procedure

The study sessions lasted 1 to 1.5hrs, beginning with a 5-minute introduction to the system. This was followed by a 15-minute walkthrough of the full system to showcase individual features. After that, the users were given 5 minutes to explore Soloist's features and interface freely. Once familiarized, they started to learn with the video lesson for about 40 minutes. The experimenter would remind users of features to use where appropriate throughout the study.

A semi-structured interview was conducted at the end of the study to obtain feedback on each feature. The interviews were recorded along with the study session and one of the authors coded and summarized the important comments. The participants were also given a questionnaire asking about individual features, the overall system, and the apparent accuracy of Soloist's algorithms, all provided on a 5-point Likert scale. Participants also rated their preference between Soloist and traditional YouTube videos. The ratings are shown in Figure 7.

### 8.4 Findings

The feedback from the participants were encouraging. Users felt excited about the demonstrated features and indicated that Soloist addresses many points of frustration with the traditional ways of learning guitar with YouTube videos.

#### 8.4.1 Overall Comments

We started by asking the participants to compare Soloist with their prior experiences of learning guitar with traditional YouTube videos. The participants unanimously preferred practicing with the Soloist system (5 strongly agreed, 3 agreed). The average ratings of the Soloist system for "enjoyable to use" and "efficient to use" were 4.5 and 4.8, respectively, while those of the traditional YouTube player were 3.9 and 3.4. P4 commented, *"This is fantastic! The interface is easy to use, and the features are very useful when learning guitar with videos."* P2 said, *"Of course it is more efficient to practice guitar using this system because I can easily find out the segments I want to practice on."* P6 requested additional time to finish the whole video as he commented, *"It really helps for practicing the solo!"* Therefore, we let him use the system for an additional 20 minutes after the main portion of the study was completed.

#### 8.4.2 Video Segmentation

The participants rated the video segmentation to be extremely useful (4.8). P5 commented, *"I usually don't listen to what the instructor says, so being able to highlight these regions is very helpful."* Multiple participants commented that they used to search for desired timestamps in a trial-and-error fashion on traditional video players while Soloist's segmentations provide clear visuals to guide the process. Some participants talked about their wish to have some annotation for the provided regions. P3 said, *"It'd be great if I can annotate the regions with texts, so I can know which part I'm not familiar*

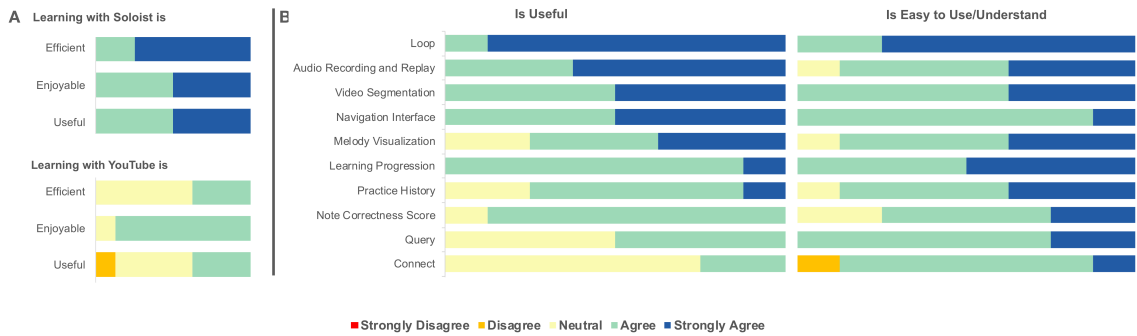

Figure 7. (A) Overall rating between Soloist and YouTube (B) Rating of individual features sorted by the usefulness scores. For each row, the left bar indicates how useful a feature is while the right bar shows how easy it is to use or understand.

*with next time I use the system."* The video segmentation accuracy had positive-leaning ratings (2 strongly agree, 3 agree, 3 neutral), showing that most participants believed the segmentations served its job. However, despite being generally positive, the mixed ratings for accuracy along with the results of our technical evaluation indicate future possibilities to improve the proposed video segmentation algorithm.

*8.4.3 Navigation Interface*

The average ratings for "easy to use" and "would be useful" for Soloist's navigation interface were 4.1 and 4.5, respectively. Notably, the participants found it useful to be able to select refined regions and loop them, evidenced by a near perfect rating (7 strongly agree, 1 agree). P3 said, *"When I use YouTube, I have to fixate my cursor on the timeline to memorize the timestamps to replay or use the arrow keys to jump back. It is kind of troublesome so it's really convenient to be able to loop these regions."* P1 highlighted the advantages of visualizing waveforms, *"Thumbnail previews of the video won't tell me where there is a sound so seeing the waveform helps."* Participants also found the separated display of voice and instrument track to be useful. P5 said, *"Yeah, it's helpful! I can easily know there is something here,"* while pointing at the waveform of the separated instrumental track. P3 emphasized its use in correcting algorithmic errors, *"It helps me easily finetune the regions when the automatic segmentation is not accurate."* This type of feedback justifies the mixed-initiative approach.

*8.4.4 In-app Recoding*

The participants all agreed or strongly agreed that Soloist's feature of rehearsal recording is useful (4.6) because they could immediately listen to what they had played. P4 elaborated on the convenience that this in-app recording affords, *"Typically, I will use my cellphone to record what I've played, but then I will have to use two devices. It's great that there's an interface integrating these functions."*

*8.4.5 Melody Visualization*

The Melody Visualization feature also received a positive-leaning rating on its usefulness (3 Strong Agrees, 3 Agree, 2 Neutral). Several insights were revealed. For instance, P1 is a novice on the electric guitar but has a degree in music. He said he mostly relies on his ear to learn music and does not require much visual information. On the other hand, P2 is a novice to both guitar and music and found it advantageous to see the visualizations of melody. He said he sometimes cannot tell the subtle differences between his playing and the reference audio. This echoes our findings in formative interviews that novices would benefit more from additional feedback notifying them of mistakes. In addition to the visualization itself, participants also highly appreciated the design of overlaying the Melody Visualization on a waveform player. P4 commented, *"With the visuals, I can select and play where I made mistakes instead of replaying the whole clip every time."*

### 8.4.6 Scoring and Progression

The participants enjoyed Soloist's gamification of guitar practice. Multiple participants (P1, P4, P6, P8) commented that the design helped track their progress and invoked their desire to color all the regions teal (*Aced*). This indicates that progression tracking helps motivate users to continue learning, which we found to be useful for novices in our formative interviews. When asked about the score offered by Soloist, P4 commented, *"Sometimes the scores are inaccurate. But that's not a big deal because I can simply modify it."* We did observe that participants sometimes found it hard to interpret a score.

### 8.4.7 Less Used Features

During the study, we found that the users mostly focused on completing the provided regions and tried *Connect* and *Query* only after the experimenter explicitly encouraged them. We hypothesize that this might be because the participants are not familiar enough with the system and that the two features are not directly relevant to their progress. For example, P4 said she thought the two features are useful but forgot about them as she was focusing on finishing the tutorial. This suggests future work to consider ways of enhancing the discoverability of features not used frequently. Retrospectively, it may also imply that for guitar solo tutorials, techniques for finding all instances of a specific query are not as useful as in other domains [24, 33], since users can loop a region multiple times to achieve similar outcomes.

### 8.4.8 Novice versus Advanced Players

We consider novices to be the primary user group as they tend to require additional feedback on technique errors and external mechanisms to reinforce their motivations for continual learning. However, our study showed that many of Soloist's features are beneficial for both novice and non-novice players, such as video navigation tools and in-app recording/replay. Moreover, though the visual feedback might not be as necessary for advanced players, P7 commented that he still enjoyed using the melody visualization to *"double-check the correctness of my self-judgment"*. As a result, we believe Soloist would also be a valuable tool to support non-novice players to learn guitar from videos.

In summary, all participants found positives in the Soloist system and showed strong preferences towards our system over traditional ways of learning guitar with videos. During the interview, participants also mentioned their own experiences in practicing instruments with videos and commented on how our system helped solve many of the pain points [41], justifying Soloist's design.

## 9 DISCUSSION AND FUTURE WORK

We discuss issues surrounding Soloist and identify potential future work on extending our system.

### 9.1 Video Navigation

Soloist offers a set of tools to navigate music videos that are primarily audio-focused. While the results from the study showed that the techniques were well-received, a limitation is that these techniques are based upon traditional mouse cursor input. While users were able to hold their guitar and control the interface simultaneously, future work can look at integrating voice- [9] or gesture-based control [2].

## 9.2 Video Segmentation

Automatic video segmentation is a crucial component of the Soloist system. Our technical evaluation shows that our method is reasonably consistent with human judgment, but discrepancies may appear since there is no single correct way to segment a video [18]. Participants of the user study also rated our segmentation method positively with an average score of 3.9 out of 5. However, we acknowledge that the number of videos we evaluated is limited and note that the labellers were authors of the paper, which may have introduced some bias in the labelling. Future work could leverage crowdsourcing to obtain a larger corpus of human-labelled data to validate our approach [18].

## 9.3 Generalizing to other Instruments

Theoretically, Soloist should be able to process videos consisting of monophonic music played by other instruments since the audio processing we use are all instrument-independent. Moreover, prior studies [37] have also found that tutorial videos share similar content formats across different instruments, strengthening our belief in Soloist's generalizability. However, we recognize that we did not formally evaluate our system with different instruments and that learners' needs may differ. Future work can investigate and contrast the learning processes of different instruments with instructional videos. Additionally, Soloist is currently limited to processing videos teaching only a single line of melody to maximize estimation accuracy. However, many types of music and instruments play multiple notes simultaneously. Therefore, future work can leverage multi-f0 detection [7] to support a wider variety of music instructional videos once the technology becomes more matured. For instruments or music that primarily consist of chords, one can use chord recognition [41] as an alternative algorithm to evaluate the practice.

## 9.4 Automatic Musical Performance Assessment

Objectively assessing musical performance is challenging [15] and perhaps impossible due to the subjective nature of music. Existing systems have tried examining different characteristics such as pitch [72] or tempo [36]. Since tutors in the videos often break down phrases and demonstrate in free rhythm, our system only calculates the correctness of pitch and lets the user identify errors in tempo and techniques by looking at the melody visualization and replaying their recordings. Our evaluation showed that the correctness score may sometimes be inaccurate and future work can investigate more robust grading systems. It can also investigate methods to automatically assess more abstract concepts in music, such as musical expressions and dynamics.

However, even if one can perfectly analyze every aspect of music and present that information to the user, it is still unclear how helpful that would be to the user. This is because self-listening, or self-assessing [27, 68], plays a crucial role in developing musicianship. Recognizing this, our system leverages a mixed-initiative approach where the system provides a score assessment but requires the user to self-listen and review the melody visualization to validate that assessment. Future work can potentially compare and contrast three different approaches to guide music learning: fully automated, mixed initiative, and human assessments.

Finally, another limitation is that Soloist relies solely on audio information, while instrument playing also requires motor skills [28, 31]. Future work can investigate the combination of audio information and motion data of the user or instrument for additional feedback.

## 9.5 Contrasting Soloist with Existing Learning Platforms and Similar Tools.

Table 2 shows a comparison between Soloist and other existing guitar learning platforms. Typical guitar learning platforms, such as YouTube and Fender Play, offer video tutorials but do not provide instantaneous feedback or specialized navigations for instrument learning. On the other hand, while Yousician provides feedback on user performance, it does not provide tutorial videos. Instead, it displays a gamified scrolling fretboard with rules similar to other music games, such as Rock Smith [76] and Guitar Hero [77].

In addition to dedicated learning platforms, web tools like YouLoop [79] or LoopTube [80] offer users the ability to loop and adjust video speed but requires manual creation of loops and do not offer feedback. Capo [78] provides waveform navigation of an imported song. Yet, it only supports audio files of songs and not tutorial videos, forgoing the crucial visual aspect of a tutor playing the guitar.

Soloist preserves the advantages of all the platforms above and further contributes a novel back-end processing pipeline enabling users to turn any online video into an interactive tutorial. However, our formative interviews also suggest that the quality of free content varies while resources of consistent quality can be found on paid platforms. Future work can investigate methods to filter out videos of lower quality and conduct longitudinal studies to compare Soloist's learning outcomes with existing music learning systems or human tutoring.

Table 2: Comparison between Soloist and other learning platforms.

|  | YouTube | Fender Play | Yousician | Soloist |
|---|---|---|---|---|
| Video Demonstrations | **Yes** | **Yes** | No | **Yes** |
| Instantaneous Feedback | No | No | **Yes** | **Yes** |
| Navigation for Guitar Learning | No | No | No | **Yes** |
| Learning Progression Tracking | No | **Yes** | **Yes** | **Yes** |
| Quality of Content | Varying | **High** | **High** | Varying |
| Scale of Content | **Large** | Limited | Limited | **Large** |

## 10 CONCLUSION

We have presented Soloist, a novel system that leverages audio processing techniques to extract musical information from music tutorial videos and turn raw videos into interactive tutorials. We have also presented a formative interview to understand the current practice of learning guitar with online videos to guide our system's design. Soloist segments videos purely based on the video's audio without requiring any additional data. Soloist also provides video navigation techniques tailored to learning with music instructional videos and utilizes a mixed-initiative design that includes both computer and human-in-the-loop to co-define the tutorial experience. We evaluated our video segmentation algorithm and showed it is reasonably consistent with human segmentation. We deployed our system on a cloud service and conducted a remote user study, which elicited unanimously positive feedback, strengthening our belief that Soloist offers a suite of techniques that can change the way people learn music with videos.

## ACKNOWLEDGMENTS

This research was supported in part by the National Sciences and Engineering Research Council of Canada (NSERC) under Grant IRCPJ 545100-18. We thank Yu-Siang Huang for insightful discussions on audio processing and Yu-Hua Chen for helping with video shooting. We also thank our colleagues at the DGP lab at the University of Toronto for their helpful feedback and the study participants.